\documentclass[11pt]{article}
\usepackage{moriond,epsfig}

\def\be{\begin{equation}}
\def\ee{\end{equation}}
\def\bea{\begin{eqnarray}}
\def\eea{\end{eqnarray}}

\begin{document}
\vspace*{4cm}
\title{THE ROLE OF LEPTON FLAVOURS IN THERMAL LEPTOGENESIS}

\author{F.X. JOSSE-MICHAUX \footnote{josse@th.u-psud.fr}}

\address{Laboratoire de Physique Theorique, Universite Paris-Sud\\
91405 Orsay Cedex, France}

\maketitle\abstracts{Thermal leptogenesis, that can be viewed as a consequence of the seesaw model, is a very natural mechanism to explain the matter anti-matter asymmetry of the Universe. Recently, lepton flavours have been included in the Boltzmann equations, modifying significantly the evaluation of the efficiency of leptogenesis to explain the observed baryon asymmetry.}

\section{Introduction}
Among the weakness of the Standard Model (SM), two at least could be explained by the fact that neutrinos are massives. Those masses, deduced from neutrino oscillation experiments, as well as the mixings between the different generation of neutrinos, are well described within the seesaw mechanism. For that purpose, at least two right handed neutrinos of Majorana type that are singlet under the SM gauge group are introduced. Their mass scale, constrained by the seesaw model, can be much higher than the electroweak scale, in the case of a natural theory, that we consider in this proceeding. Leptogenesis, in the thermal scenario, is the creation of a net lepton number in the very early Universe from the decay of heavy right-handed neutrinos, lepton asymmetry that is (partially) converted into a baryonic one through sphalerons processes. 
\subsection{The baryon asymmetry of the Universe}
The  baryon asymmetry, that is the difference between the number density of baryon and anti-baryon normalised to the number density of photons, is deduced from cosmological observation\cite{WMAP} to be:
\begin{equation}
\eta_{\cal B}=\frac{n_{b}-n_{\bar{b}}}{n_{\gamma}}=(6.1\pm0.2)\times10^{-10}.
\end{equation}
To explain this asymmetry in a inflationnary universe, three conditions given by Sakharov\cite{Sak} are necessary: baryon number violation, C and CP violation, and departure from thermal equilibrium. These conditions are satisfied in the Standard Model, but hardly enough to explain the amount of baryon asymmetry. One needs to go beyond the SM. 
\subsection{The Seesaw mechanism}
In the seesaw model \cite{seesaw}, neutrinos of Majorana type are added to the SM particule content to explain the smallness of the observed masses. The extended Lagrangian, in the mass basis of the right-handed neutrino $N$ and of the charged lepton $\ell$, reads:
\begin{equation} \label{lag}
{\cal{L}}={\cal L}_{SM}-\frac{1}{2}\bar{\hbox{\ }N_i}M_{N_i}N_{i}-h_{\nu}^{\alpha j} \bar{\ell_{\alpha}} N_{j}\phi \ .
\end{equation}
$M_{N_i}$ is a mass matrix, diagonal in the basis we choose, and $h_{\nu}^{\alpha i}$ is a complex Yukawa coupling. The diagonalisation of the neutrino mass matrix gives two mass eigenstate (per generation): one with a mass $\simeq M$, and the other with the mass $m_{\nu}\simeq v^2 h_{\nu} M^{-1} h_{\nu}^{T}$, where $v$ is the vacuum expectation value of the Higgs field $\phi$. For  light neutrino mass of about 1 eV, and natural couplings $h_{\nu}$, $M$ should be around $10^{9} GeV$ or more.

\section{Thermal Leptogenesis}
In the thermal leptogenesis scenario \cite{lepto0}, the right handed neutrino $N$ are produced in the thermal bath by scattering processes that occur at a temperature $T\simeq M_{N}$. These $N$ decay into lepton (plus higgs) and anti-lepton (plus higgs$^{*}$), and if the decay violate CP a net lepton number is produced. Depending on the strength of inverse reaction, that is, depending on if inverse reaction are out-of-equilibrium or not, an asymmetry will survive, and will be partly converted into a baryon asymmetry by the ${\cal B+L}$ - fast violating processes that are in equilibrium above the electroweak phase transition.
\subsection{The standard picture: the one-flavour approximation}
As in seesaw models leptogenesis qualitatively occurs, the first goal was to show that leptogenesis can quantitatively account for the observed matter-antimatter asymmetry. This has been done in the so called one-flavour approximation, where the lepton asymmetry is produced in one dominant flavour from the decay of the lightest right-handed neutrino, assuming a strong hierarchy:  $M_{N_1}\ll M_{N_2},M_{N_3}$. In this picture, it has been shown that leptogenesis can quantitatively works \cite{lepto1,lepto2}. The baryon asymmetry is estimated to be: 
\begin{equation}
Y_{\cal B}=\frac{n_{b}-n_{\bar{b}}}{s}\simeq a_{sph}\times Y_{N_1}^{eq}(T\gg M_{N_1})\times \epsilon_{N_1} \times \eta \ ,
\end{equation}
where $a_{sph}\sim1/3$ is the fraction of lepton asymmetry converted into a baryonic one through the sphaleron interactions, $Y_{N_1}^{eq}(T\gg M_{N_1})\simeq 4\times 10^{-3}$ is the equilibrium number density of $N_1$ at the beginning of the leptogenesis era after the reheating period. The CP violation in the decay is parametrized by the CP asymmetry $\epsilon_{N_1}$, which is defined as: 
\begin{equation}
\epsilon_{N_1}=\frac{\Gamma(N_1 \rightarrow H \ell)-\Gamma(\bar{N}_1 \rightarrow \bar{H} \bar{\ell})}{\Gamma(N_1 \rightarrow H \ell)+\Gamma(\bar{N}_1 \rightarrow \bar{H}\bar{\ell})} \ .
\end{equation}
The last factor, $\eta$, is the efficiency of the process, and highlights the competition between the production of a lepton asymmetry by decay and its destruction by inverse reaction (inverse decay and scattering processes). If the inverse reaction are fast compare to the Hubble expansion rate at the temperature $T\simeq M_{N_1}$, then the lepton asymmetry will be strongly wash-out and not enough baryon asymmetry will be created.
\subsection{The lepton flavour}
The mass eigenstates of the particule contributing to the Boltzmann equations (BE) are determined by the reactions which are fast compare to the processes included in the BE. But at the temperature leptogenesis occurs, $T_{lepto}\simeq M_{N_1} \simeq 10^{9}$ GeV, as the fields acquire large thermal mass, the interaction involving charged Yukawa couplings  develop thermal corrections:
\begin{equation}
\Gamma_{\ell_{\alpha}} \simeq 5\times10^{-3}\times h_{\ell_{\alpha}}^{2}\times T \ .
\end{equation}
Depending on the Yukawa couplings $h_{\ell_{\alpha}}$, the interactions can be faster than the Hubble expansion rate at $T_{lepto}$, and have to be taken into account in the calculation of the proper mass eigenstates. More precisely, as $T_{lepto}\simeq M_{N_1}$, if $M_{N_1}$ is above about $10^{12}$ GeV, none of the interaction that bear the flavour information is in equilibrium, thus making indistinguishable the different flavour: the one-flavour approximation is valid. But if $M_{N_1}$ is below $10^{12}$ GeV  the tau-Yukawa interactions are in equilibrium and two flavour are distinguishable: the flavour $\tau$ and an orthogonal flavour compose of $\mu$ and $e$.Iif $ M_{N_{1}} \leq 10^{9} $ GeV the muon-Yukawas are in equilibrium too and the lepton asymmetry is projected onto a three flavour-space $\tau$, $\mu$, $e$.
\section{Flavoured leptogenesis}
In thermal leptogenesis  the constraint that the reheating temperature should be above than $M_{N_1}$ in order not to wash-out the produced baryon asymmetry, but $T_{rh}$ should also not being too big, in order to avoid overclosure problem. Therefore, the lowest $M_{N_1}$ is the preferred choice. On the other hand, a lower bound on $M_{N_1}$ has been derived in the one flavour approximation \cite{DI}, $M_{N_1} \geq 10^{9}$ GeV (in the case where $N_1$ is produced by thermal scatterings). As explained before, the flavour content should be taken into account: flavour matters \cite{flav}. We thus have to define a CP asymmetry for each (distinguishable) flavour, 
\begin{equation}
\epsilon_{N_1,\ell_{\alpha}}=\frac{\Gamma(N_1 \rightarrow H \ell_{\alpha})-\Gamma(\bar{N}_1 \rightarrow \bar{H} \bar{\ell_{\alpha}})}{\Gamma(N_1 \rightarrow H \ell_{\alpha})+\Gamma(\bar{N}_1 \rightarrow \bar{H}\bar{\ell_{\alpha}})} \ ,
\end{equation}, as well as individual efficiencies $\eta_{\alpha}$, so that the baryon asymmetry, when flavours are accounted for, reads: 
\begin{equation}
Y_{\cal B}\simeq a_{sph}\times Y_{N_1}^{eq}(T\gg M_{N_1})\times \sum_{\alpha}\epsilon_{N_1 \ell_{\alpha}} \times \eta_{\: \ell_{\alpha}} \ .
\end{equation}
Recall that in the one-flavour approximation we have $\sum_{\alpha}\epsilon_{N_1 \ell_{\alpha}} \times \eta_{ \ell}$, where $\eta_{\ell}$ is  the efficiency factor for the total lepton asymmetry. This comes from the fact that in the Boltzmann equations for the number densities, each (distinguishable) flavour is washed-out with a strength $\propto \tilde{m_{\alpha}}/m^{*}$, where $\tilde{m_{\alpha}}$ is the rescaled partial decay width $\Gamma(N_1 \rightarrow H \ell_{\alpha})$ and $m^{*}\simeq 10^{-3} $eV is the ``equilibrium neutrino mass", the rescaled Hubble expansion rate at $T_{lepto}$. In the one-flavour case, the total lepton asymmetry is washed-out with a strength $\propto \sum_{\alpha} \tilde{m_{\alpha}}/m^{*}$, then possible flavour mis-alignment can enhanced the amount of lepton asymmetry. Indeed, the efficiency is maximum for $\tilde{m}\simeq m^{*}$, but the mass infered from neutrino oscillations, $m_{atm}\simeq 5\times10^{-2}$eV and $m_{sol}\simeq9\times10^{-3}$eV are both above $m^{*}$, and thus the region $\tilde{m}\geq m^{*}$ is more attractive, even if less efficient for leptogenesis. Including flavour, one can have $\sum_{\alpha} \tilde{m_{\alpha}} \geq m^{*}$, even if one of the flavour is weakly washed-out $\tilde{m_{\alpha}}\simeq m^{*}$. The efficiency for this flavour will be (close to) maximum, and the flavour will dominate the lepton asymmetry (unless its CP asymmetry $\epsilon_{N_{1}, \ell_{\alpha}}$ is too small...), allowing for a sufficient amount of baryon asymmetry, even if the total wash-out is strong $\sum_{\alpha} \tilde{m_{\alpha}} \gg m^{*}$.

Another feature of the inclusion of lepton flavour concern the upper bound on the light neutrino mass scale. In the one-flavour case, an upper bound of $0.15$ eV was derived \cite{mass} on the light neutrino mass scale, from the recquirement that the total wash-out should not be too strong. But as we have seen, flavour mis-alignemt allows successfull leptogenesis even for a strong total wash-out if one flavour is in weak or mild wash-out $\tilde{m_{\alpha}}\simeq m^{*}$, and therefore no upper bound on $m_{\nu}$ holds from leptogenesis: the cosmological bound is saturated.
\section{Conclusion}
In seesaw models leptogenesis qualitatively occurs: a right handed neutrino produce a lepton asymmetry via out-of-equilibrium CP violating decays. It has been shown that it quantitatively explain the observed amount of baryon asymmetry in the one-flavour approximation. Including lepton flavours, the computation of the baryon asymmetry is modified, and some constraint are relaxed, from possible mis-alignment of the flavours. For the good range of temperature, flavoured treatment of leptogenesis is more accurate.
\section*{Acknowledgments} 
I would like to thanks the organising committee for inviting me at this pleasant conference. This work was funded in part by the MCA grant number 516636.

\end{document}